\documentstyle[12pt]{article}

\begin{document}
\begin{titlepage}

\begin{center}
{\hbox to\hsize{\hfill April 2007 }}

\bigskip \vspace{3\baselineskip}

{\Large \bf Electroweak Higgs as a pseudo-Goldstone boson \\ of broken scale invariance \\ }

\vspace{2 cm}

{\bf Robert Foot, Archil Kobakhidze and Raymond R. Volkas \footnote{E-mail: foot@physics.unimelb.edu.au, archilk@physics.unimelb.edu.au, r.volkas@physics.unimelb.edu.au}\\}

\smallskip

{ \small \it School of Physics, Research Centre for High Energy Physics,\\ 
The University of Melbourne, Victoria 3010, Australia\\}

\bigskip \vspace*{.5cm}

{\bf Abstract}\\

\end{center}

\noindent 

{\small We point out that it is possible to associate the electroweak Higgs boson with the pseudo-Goldstone boson of broken scale invariance, thus resolving the hierarchy problem in a technically natural way. We illustrate this idea with two specific gauge models. Besides 
being consistent with all currently available experimental data, 
both models maintain the predictive power of the standard model, since the first model 
has only one additional parameter beyond the standard model, and the 
second has the same number of free parameters as the standard model.}

\bigskip

\bigskip

\end{titlepage}

\baselineskip=16pt

\section{Introduction}

Understanding the origin of mass is one of the key problems within 
the standard model (SM). The chiral nature of the gauge symmetry of 
the Standard Model forbids masses for quarks and leptons, apart from 
right-handed neutrino Majorana masses. Taking neutrinos to be Dirac, the 
only mass parameter in the theory is the electroweak $\mu^2$ mass 
parameter in the Higgs potential.   
Setting it to zero renders the SM scale 
invariant at the classical level. However, the scale invariance is  
known to be anomalous; it is broken at the quantum level. This manifests 
through the important effect called dimensional transmutation. In 
particular, in a scale invariant theory a scalar field can develop 
a vacuum expectation value (VEV) radiatively as a result of the quantum 
conformal anomaly \cite{Coleman:1973jx}. On general grounds one can argue that the existence 
of a light scalar field, with its mass generated entirely at loop level, is 
inevitable within scale invariant models that involve also 
multiple scalar fields \cite{Gildener:1976ih}. This light 
scalar field is nothing but a pseudo-Goldstone boson (PGB) accompanying 
the breaking of the anomalous scale invariance. It is tempting 
to associate this PGB with the electroweak Higgs boson. The reasons are 
obvious. Firstly, a massless theory is more predictive than the 
corresponding theory with {\it a priori} unknown mass parameters. Secondly, 
and perhaps more importantly, the electroweak scale generated through 
the dimensional transmutation will be 
radiatively stable, thus resolving the technical aspect of the hierarchy  
problem. Unfortunately, this very appealing theoretical 
framework applied within the SM 
predicts a very light Higgs boson with $m_{\rm h} < 10$ GeV and also requires 
the top quark to be light: $m_{\rm t} \stackrel{<}{\sim} 40\ GeV$.  
Both of these predictions are in sharp contradiction with observations.

A quick inspection of the Coleman-Weinberg effective potential reveals the way to 
circumvent the above problem in weakly-coupled (perturbative) theories. 
One needs to ensure the overall dominance of the bosonic contributions 
to the effective potential over the fermionic ones, such that the 
experimental lower bound on the Higgs boson mass is satisfied while 
keeping the coupling constant in the perturbative domain.  
The simplest way to achieve this is by simply adding one (or more) scalar 
fields to the theory.\footnote{Scale invariant models with additional scalars  
have been considered in Refs. \cite{Hempfling:1996ht} and \cite{meiss} (see also the discussion of the  scale-invariant case in Ref. \cite{Espinosa:2007qk}). The philosophy of those models was somewhat different though, 
with the additional scalar gaining a large VEV in both models 
which means that the electroweak Higgs could not be interpreted as the PGB of broken scale invariance. Also in Ref. \cite{Hempfling:1996ht}  
there was an additional $U(1)_X$ gauge boson. 
The models which we consider are simpler, with less parameters than  
these alternative models.} 
The minimal model of this type involves a Higgs potential 
with just three parameters -- which is only one more parameter than the  
SM case. We study this model in section II, where we show 
that it does provide a phenomenological consistent  
theoretical framework to realise the Higgs boson as a PGB of broken 
scale invariance.

Extended scalar sectors are a feature of many theories beyond the standard model. 
One such theory, with a particularly simple Higgs sector, is the mirror model \cite{mirror}.   
In that theory, one essentially has two isomorphic sectors of  
particles, the standard particles and a `mirror' sector. The mirror  
particles are governed by a Lagrangian of exactly the same form 
as the standard model, so that a discrete $Z_2$ mirror symmetry can be  
defined interchanging the ordinary and the mirror particles. 
If we make the theory scale invariant, by eliminating the $\mu^2$ mass 
parameter in the Higgs potential, then we can generate 
electroweak symmetry breaking radiatively, via the Coleman-Weinberg mechanism. 
The Higgs potential then has just two parameters (the same as in the 
SM) since the $Z_2$ mirror symmetry in the model fixes the 
quartic coupling constant in the mirror sector to be the same as the corresponding
coupling constant in the ordinary particle sector. 
We show in section III that this classically scale  
invariant model  
is phenomenologically consistent, and also realises the Higgs boson 
as a PGB of spontaneously broken scale invariance.

\section{The next-to-minimal scale-invariant Standard Model}

Consider a minimal extension of the scale-invariant SM with a single extra 
real scalar field. The most general renormalisable potential is
\begin{equation}
V_0(\phi,S) = \frac{\lambda_1}{2}(\phi^{\dagger}\phi)^2 + \frac{\lambda_2}{8}S^4
+ \frac{\lambda_3}{2}(\phi^{\dagger}\phi)S^2~,
\label{a1}
\end{equation}
where $\phi$ is the electroweak Higgs doublet and $S$ is the real singlet field.  
observe that the above potential (as well as total Lagrangian) has an accidental
discrete $Z_2$ symmetry: $S\to -S$. We parameterise the fields (in the unitary
gauge) as:
\begin{equation}
\phi= {r \over \sqrt{2}}
\bordermatrix{&\cr                &0 \cr                & \cos\omega}~,~~S=r\sin\omega
\label{a2}
\end{equation}
In this parameterisation, the potential (\ref{a1}) takes the form
\begin{equation}
V_0(r, \omega)=r^4\left(\frac{\lambda_1}{8}\cos^4\omega
+\frac{\lambda_2}{8}\sin^4\omega+\frac{\lambda_3}{4}\sin^2\omega\cos^2\omega  \right).
\label{a3}
\end{equation}
Note that the radial component $r$ of the Higgs fields (\ref{a2})
factors out in the absence of the tree-level mass parameter. 
The renormalised quantum-corrected potential, in addition to (\ref{a3}), includes 
a sum of $\delta V_{k-{\rm loop}}$ contributions generated at $k$-loop 
level ($k=1,2,\ldots$). Each $k$-loop contribution is a $k^{\rm th}$ order
polynomial in $\log(\phi/\Lambda)$, where $\Lambda$ is some
renormalisation scale. Perturbation theory is valid if the logarithms are not
too large, so that $V_{0}>\delta V_{\rm 1-loop}>\delta V_{\rm 2-loop}
> \ldots$ is satisfied. In general, the minimisation of even a 1-loop corrected
effective potential involving multiple scalars cannot be done analytically, but one
may resort to a numerical analysis. Here we instead follow the approximate
analytic method suggested in \cite{Gildener:1976ih} which is suitable in weakly coupled
scale invariant theories.  

Following \cite{Gildener:1976ih}, we first ignore perturbatively small radiative
corrections and concentrate on the tree-level potential (\ref{a3}). 
Minimising the potential, assuming that $r \neq 0$ but is otherwise
at this stage arbitrary, leads to two possible cases (we ignore
a third case with unbroken electroweak symmetry because it 
is phenomenologically not viable). If $\lambda_3 > 0$ then
\begin{equation}
\langle \sin\omega \rangle =0~,~ \langle r \rangle = \sqrt{2} \langle \phi \rangle \equiv 
v\approx 246~ {\rm GeV}~,~<S>=0~,
\label{a5}
\end{equation}
with
\begin{equation}
\lambda_1(\Lambda)=0~.
\label{a55}
\end{equation}
In this case only the electroweak Higgs develops a nonzero VEV. If $\lambda_3 < 0$ then  
\begin{eqnarray}
\langle \tan^2\omega \rangle &=& \epsilon~, \nonumber \\
~\sqrt{2}\langle \phi \rangle  &=& 
\langle r \rangle \left(\frac{1}{1+\epsilon}\right)^{1/2}\equiv v\approx 246~{\rm GeV}~,~
\langle S \rangle = v\langle \tan\omega \rangle ~,
\label{a6}
\end{eqnarray}
with
\begin{equation}
\lambda_3(\Lambda)+\sqrt{\lambda_1(\Lambda)\lambda_2(\Lambda)}=0
\label{a66}
\end{equation}
where $\epsilon \equiv \sqrt{\frac{\lambda_1(\Lambda)}{\lambda_2(\Lambda)}}$. In this case 
both scalar fields develop VEVs and the
discrete $Z_2$ symmetry is also broken spontaneously.\footnote{This might cause a problem
with cosmological domain walls unless $\langle S \rangle$ is sufficiently small. If it
is very small but nonzero, then a network of domain walls would form with an equation of
state ($p = w\rho$) parameter  $w=-2/3$ in the non-relativistic limit.  This is somewhat interesting for
dark energy reasons, though present indications are that a $w$ value closer to $-1$ is
preferred by the data.}

Relations such as Eqs.(\ref{a55}) and (\ref{a66}) can be satisfied by an
appropriate choice of the renormalizsation point $\mu=\Lambda$ \cite{Gildener:1976ih},
where the running coupling constants depend on $\mu$, and $\Lambda$
is the specific value where the required relations hold.  In
each case, the tree-level potential then has a flat direction along
the vacuum solution, and the relation Eq.(\ref{a55})  [or Eq.(\ref{a66})] removes a dimensionless parameter in favour of the 
renormalisation point $\Lambda$ which has the dimension of mass (dimensional 
transmutation). Because the tree-level potential vanishes along a specific 
direction, the $1$-loop correction to it will dominate along that direction.

Next we calculate the tree-level masses by expanding the Higgs potential, Eq.(\ref{a1}),   
around the vacuum: $\phi = \langle \phi \rangle + \phi'$, $S = \langle S \rangle + S'$. 
In each case there are two physical scalars, but 
only one of these gains mass at tree-level, since there is a flat direction  
in the Higgs potential. Let us call $H$ the state that gets mass at tree-level, 
and $h$ the state which is massless at tree-level (the PGB of broken scale invariance). 
For the first case, where $\lambda_3 > 0 \ \Rightarrow \langle S \rangle = 0$ [Eq.(\ref{a5})],
we find
\begin{eqnarray}
m_H^2 = {\lambda_3 v^2 \over 2} \ , \ \ H = S.
\end{eqnarray}
The PGB in this case is $h = \phi'_0$.

In the second case, where $\lambda_3 < 0$ and both $S$ and $\phi$ gain VEVs, we find
\begin{eqnarray}
m_H^2 = \lambda_1 v^2 - \lambda_3 v^2 \ , \ \ 
H = -\sin\omega \phi'_0 + \cos\omega S'.
\end{eqnarray}
In this case the PGB is $h = \cos\omega \phi'_0 + \sin\omega S'$.

Let us now calculate the mass of the PGB boson for each pattern of symmetry 
breaking.  The $1$-loop correction to the tree-level 
potential (\ref{a3}) (along the radial direction) has the form \cite{Gildener:1976ih}
\begin{equation}
\delta V_{\rm 1-loop}= Ar^4 \ + \
Br^4\log\left(\frac{r^2}{\Lambda^2}\right )~,
\label{14}
\end{equation}
where
\begin{equation}
A=\frac{1}{64\pi^2 \langle r \rangle^4}\left[3{\rm
Tr}\left(M_V^4\log\left(\frac{M_V^2}{\langle r \rangle^2}\right)\right)+{\rm
Tr}\left(M_S^4\log\left(\frac{M_S^2}{\langle r \rangle^2}\right)\right)-4{\rm
Tr}\left(M_F^4\log\left(\frac{M_F^2}{\langle r \rangle^2}\right)\right) \right ]~,
\label{15}
\end{equation}
and 
\begin{equation}
B=\frac{1}{64\pi^2 \langle r \rangle^4}\left [3{\rm Tr}M_V^4+{\rm Tr}M_S^4-4{\rm
Tr}M_F^4
\right ]~.
\label{16}
\end{equation}
The traces in the above equations go over all internal degrees of
freedom, and $M_{V,S,F}$ are the tree-level masses respectively for
vectors, scalars and fermions evaluated for the given VEV pattern.

The stationary condition $\frac{\partial \delta V_{\rm 1-loop}}{\partial
r}|_{r=\langle r \rangle}=0$  implies the relation 
\begin{equation}
\log\left(\frac{ \langle r \rangle}{\Lambda} \right)=-\frac{1}{4}-\frac{A}{2B}.
\label{17}
\end{equation}
The PGB mass can be calculated directly from $\delta V_{\rm 1-loop}$.
Using (\ref{17}) one finds \cite{Gildener:1976ih}:
\begin{eqnarray}
m_h^2  & = & \frac{\partial^2 \delta V_{\rm 1-loop}}{\partial r^2 }
|_{r = \langle r \rangle} = 8 B \langle r \rangle^2  \nonumber \\
& = & 
\frac{1}{8\pi^2 \langle r \rangle^2}\left [3{\rm Tr}M_V^4+{\rm Tr}M_S^4-4{\rm
Tr}M_F^4
\right ]~.
\label{18}
\end{eqnarray}
Applying this equation to determine the mass of the PGB, we find:
\begin{eqnarray}
m_h^2 &=& {1 \over 8\pi^2 \langle r \rangle^2} \left[ 6m_W^4 + 3m_Z^4 + m_H^4 - 12 m_t^4\right]
\nonumber \\
& \approx & \frac{m_H^4 \cos^2\omega}{8\pi^2v^2}~,
\label{a9}
\end{eqnarray}
since we need  $m_H$ to dominate 
over the other terms if the PGB mass is to be larger than the experimental 
lower limit of about $\sim 115$ GeV.

Precision electroweak tests put an upper bound on the Higgs boson mass. The current upper limit for the   
standard model Higgs is  $m_{higgs} < M_{EW}$, with $M_{EW} \approx 186$ GeV at 95\%  C.L. \cite{pdg}. Since the Higgs boson mass gives a radiative correction at l-loop level via a log term,  we can find the corresponding limit  in this model by the replacement 
$\lambda_{SM}^2 \log m_{higgs}^2 \to \lambda_{SM}^2 \cos^2 \omega \log m^2_h + \lambda_{SM}^2 \sin^2 \omega \log m^2_H$. Thus, the limit on the scale invariant model from precision  
electroweak tests is
\begin{equation}
\left(\frac{m_h}{m_H}\right)^{c_{\omega}^2}m_H<M_{EW}~,
\label{a10}
\end{equation}  
where $c_\omega \equiv \cos\omega$. Clearly, in the first symmetry breaking scenario, where $\cos\omega = 1$, this bound only constrains the
mass of the PGB $h$ to be less than $M_{EW}$ which can easily be satisfied (as is also the case
in the  SM).
In the second case, where $\cos\omega$ is essentially a free parameter, the above 
bound gives a contraint on $\omega$ (given the relation, Eq.\ref{a9}). 
Using the current experimental bound, $M_{EW}\approx 186$ GeV, we  
find that $\tan\omega <0.65$. That is to say, the PGB mainly ``resides'' in the 
electroweak doublet.

Note that the above analysis with one real scalar field can be simply extended to $N$ real  scalar fields.
Taking for simplicity an $O(N)$ symmetric potential, we simply need to replace 
$S^2 \to \sum^N_{i=1} S_i^2$ in the potential, Eq.(\ref{a1}).  
In this case only the $\lambda_3 > 0$ region, where $\langle S_i \rangle = 0$,  
is phenomenologically viable, since having $\langle S_i \rangle \neq 0$ would 
lead to massless Goldstone bosons from the spontaneous breaking of $O(N)$ symmetry. 
Also, if $\langle S_i \rangle = 0$ we can also give $S$ gauge quantum numbers. 
For example, having $S$ complex and transforming as an $SU(3)_c$ colour triplet would be  
equivalent to having $N = 6$ real scalar fields with an $O(6)$ symmetric potential.  
Having $N$ scalar fields, will give a factor $N$ in the right-hand side of Eq.(\ref{a9}) and 
thus reduce the mass of the heavy scalar (for a fixed $m_h$).

To find the domain of validity of perturbation theory we need to look at the renormalisation group equations \cite{Cheng:1973nv}:
\begin{eqnarray}
4\pi^2{d \lambda_1 \over dt} &=& 3\lambda_1^2 \ +\ {\lambda_3^2 \over 4} N       \nonumber \\
4\pi^2 {d \lambda_2 \over dt} &=& \left[{N+8 \over 4}\right] \lambda_2^2 
\ + \ \lambda_3^2      \nonumber \\
4\pi^2 {d \lambda_3 \over dt} &=& {3 \lambda_1 \lambda_3 \over 2}\ + \ 
{(N+2) \over 4} \lambda_2 \lambda_3
\ + \ {\lambda_3^2 \over 2},
\end{eqnarray}
where $t=\log\mu$.
Contributions from gauge and Yukawa coupling constants can be approximately  
neglected relative to the Higgs potential couplings constants. 

These equations must be supplemented by the constraint Eq.(\ref{a55}) or (\ref{a66}) 
for the two symmetry breaking scenarios of interest. Due to the relatively  
large $m_H$ mass, the position of the Landau pole, $\mu_L$, is typically only a few orders of 
magnitude above the weak scale in both symmetry breaking scenarios. 
For example, for $m_h$ at the experimental limit of 115 GeV, taking $N=1$ ($N=6$) 
and the case where $\lambda_3 > 0$ so that $\langle S \rangle = 0$, 
we find, numerically solving the equations, that 
$\mu_L \approx 20 \Lambda \approx 10^4$ GeV  
($\mu_L \approx 4 \times 10^2 \Lambda \approx 10^5$ GeV).  [Note that $\Lambda$ is determined from 
Eq.(\ref{17})].

\section{Electroweak Higgs as a PGB of broken scale invariance in mirror models} 

In the previous section we showed that extending the Higgs sector in a scale invariant 
theory allows consistent models giving a naturally light and radiatively stable 
Higgs boson. Many theories beyond the SM actually require an extended 
Higgs sector. One theory with a particularly simple Higgs potential is the mirror 
matter model \cite{mirror}.  In the simplest version of that theory each type of ordinary particle
(other than the graviton) has a distinct mirror partner. The ordinary and mirror
particles form parallel sectors, each with 
gauge symmetry $G_{\rm SM} = SU(3) \otimes SU(2) \otimes U(1)$, so that the overall gauge group is 
$G_{\rm SM}\otimes G_{\rm SM}$. The interactions within each sector 
are governed by Lagrangians of exactly the same form, except that mirror 
weak interactions are right-handed.\footnote{There is a related model  
where the interchange symmetry is just an internal $Z_2$, unrelated to parity. 
In that version, the ``shadow'' weak interactions are left-handed.} 
In other words, the full Lagrangian 
has the form 
\begin{equation} 
{\cal L} = {\cal L}_{1} + {\cal L}_{2}+{\cal L}_{\rm mix}~, 
\label{1} 
\end{equation}
where ${\cal L}_1$ is the usual Lagrangian of the SM, 
while ${\cal L}_2$ is the Lagrangian for the mirror SM. They 
are related by a parity symmetry $\cal P$, such that ${\cal P} {\cal L}_1{\cal 
P}^{-1}={\cal L}_2$, which is imposed as an exact symmetry of the 
whole Lagrangian: ${\cal P}{\cal L}{\cal P}^{-1}={\cal L}$.  
The third term ${\cal 
L}_{\rm mix}$ contains only two parity invariant renormalisable 
interactions that couple the ordinary sector with the mirror one:
\begin{equation}
{\cal L}_{\rm mix}= \epsilon F_{\mu\nu}^{1}F^{2~\mu\nu}+2\lambda
(\phi_1^{\dagger}\phi_1)(\phi_2^{\dagger}\phi_2)~,
\label{2} 
\end{equation}   
where $F_{\mu\nu}^{1}$ and $F_{\mu\nu}^{2}$ are the $U(1)$ field strengths 
for the ordinary and mirror sectors and $\phi_1$ and $\phi_2$ are 
ordinary and mirror electroweak Higgs doublets, respectively.  Thus we 
have only two extra parameters, $\epsilon$ and $\lambda$, in addition to 
those of the ordinary SM. The $U(1)$ kinetic mixing term 
will not be of interest to us here. 

The most general tree-level Higgs potential of the scale invariant
mirror model can be expressed as 
\begin{equation}
V_0(\phi_i)=\lambda \left( \phi_1^{\dagger}\phi_1 +
\phi_2^{\dagger}\phi_2  \right)^2 +
\delta \left( \phi_1^{\dagger}\phi_1 \phi_1^{\dagger}\phi_1 +
\phi_2^{\dagger}\phi_2 \phi_2^{\dagger}\phi_2 \right) \ .
\label{3}
\end{equation}
This potential is bounded from below if $\lambda+\delta \ge 0$ and
$\lambda + {\delta \over 2} \ge 0$. We take the Higgs field $\phi_i$ of each
sector in unitary gauge and express them as
\begin{equation}
\phi_1=\frac{r}{\sqrt{2}}
\bordermatrix{&\cr                &0 \cr                & \cos\omega}~,
\ \phi_2 = \frac{r}{\sqrt{2}} 
\bordermatrix{&\cr                &0 \cr                & \sin\omega}~.
\label{4}
\end{equation}
Then the potential (\ref{3}) takes the form,
\begin{equation}
V_{0}(r, \omega)=\frac{r^4}{4}\left[\lambda + \delta (\cos^4 \omega +
\sin^4 \omega)\right] \ . 
\label{5}
\end{equation}
As in the previous section, we can minimise the potential, assuming $r \neq 0$, which leads to 
two cases (depending on the sign of $\lambda$):
\begin{eqnarray}
{\rm Broken-P \ case \ for} ~\lambda > 0 &:&\quad  \sin\omega  = 0\ \ {\rm or}\ \ \cos\omega = 0\quad 
{\rm requiring}\quad  \lambda + \delta = 0; \label{8}
\nonumber \\
{\rm Unbroken-P \ case\ for }~ \lambda < 0 &:&\quad \sin\omega  = \cos\omega = {1 \over \sqrt{2}}\quad 
{\rm requiring}\quad \lambda + {\delta \over 2} = 0.
\label{7}
\end{eqnarray}
As before, the relations between $\lambda$ and $\delta$ will be satisfied by an
appropriate choice of the renormalisation point $\mu=\Lambda$.
For the broken-P case, we have either $\langle\phi_1\rangle$ or $\langle\phi_2\rangle$ 
being zero.  Since we are identifying sector $1$ as the ordinary sector, 
we shall concentrate on the $\sin\omega = 0$ configuration.

Next we calculate tree-level mass-squared matrix at $\mu = \Lambda$:
\begin{eqnarray}
{\rm Broken-P \ Case}&:& M^2_{ij} = \lambda \langle r \rangle^2 \left[\begin{array}{cc}
0 & 0 \\
0 & 1 \end{array}\right]
\nonumber \\
{\rm Unbroken-P \ case}&:& M^{2}_{ij} = \lambda \langle r \rangle^2\left[\begin{array}{cc}
-1 & 1 \\
1 & -1 \end{array}
\right] 
\label{9xx}
\end{eqnarray}
The above mass matrices each have vanishing 
determinant, so one of the physical scalar fields, the PGB,
is massless at tree level. As usual, the conformal anomaly means it obtains a relatively 
small mass from the $1$-loop correction to $V_{0}$. 

Also, it can be easily seen that the matrices in Eq.(\ref{9xx}) have 
non-negative eigenvalues provided that $\lambda > 0$ ($\lambda < 0$) 
for the broken- (unbroken-) P cases. 
Thus, there are only two physically distinct minima of the 
potential (\ref{5}):  
\begin{equation} 
\langle r \rangle \equiv v \approx 246~ {\rm GeV},\quad \langle\cos\omega\rangle = 
1,\quad {\rm 
if}~\lambda>0~,  
\label{10} 
 \end{equation}   
and 
\begin{equation}
\langle r \rangle \equiv \sqrt{2} v = \sqrt{2}\cdot 246~{\rm
GeV},\quad \langle\cos\omega\rangle = \frac{1}{\sqrt{2}},\quad {\rm if}~\lambda<0~.
\label{11}
\end{equation}

In case (\ref{10}) the electroweak symmetry is broken in the SM 
sector, while the electroweak symmetry in the mirror sectors is 
intact\footnote{$SU(2)\otimes U(1)$ electroweak symmetry in the mirror 
sector is eventually broken through mirror $SU(3)$ quark condensation. 
For details, see Ref. \cite{broken}.}, and hence the $Z_2$ parity symmetry is 
spontaneously broken.  In this model, the scalar sector consists of the standard Higgs boson, 
which is massless at tree level, and a complex mirror-doublet of bosons with 
mass squared: $m_H^2 = \lambda v^2 \approx \lambda (246)^2\  {\rm 
GeV^2}$.  

In case (\ref{11}), the $SU(2)\times U(1)$ symmetry is broken in each sector 
so the $Z_2$ mirror symmetry remains exact. 
Consequently, the gauge bosons and fermions in each sector obtain the same 
masses as the corresponding particles in 
the SM.  At tree level there is one massive scalar, with  
$m_H^2 = -4\lambda v^2 \approx -4\lambda (246) \  {\rm GeV^2}$, and one 
massless state. This massless 
state corresponds to the PGB of broken scale invariance.  The mass eigenstates 
are parity eigenstates, maximal superpositions of the ordinary and mirror  
physical Higgs bosons.

Let us now calculate the mass of the PGB boson for each pattern of symmetry breaking. 
As briefly reviewed in section II, the mass of the PGB is, in general, given by Eq.(\ref{18}). 
Applying this to the spontaneously broken mirror symmetry case 
(\ref{10}) we obtain, 
\begin{eqnarray}
m_h^2 &\simeq & {1 \over 8\pi^2 v^2} \left[ 6m_W^4 + 3m_Z^4 + 4m_H^4 -
12m_t^4 \right]
\nonumber \\
& \approx &
{m_H^4 \over 2\pi^2 v^2}  \ \ {\rm for }\ m_h \stackrel{>}{\sim} 115\
{\rm GeV}, 
\end{eqnarray}
while for the unbroken mirror symmetry case (\ref{11}) we find:
\begin{eqnarray}
m_h^2 &\simeq & {1 \over 16\pi^2 v^2} \left[ 12m_W^4 + 6 m_Z^4 + m_H^4 -
24 m_t^4 \right]
\nonumber \\
& \approx &
{m_H^4 \over 16\pi^2 v^2}  \ \ {\rm for }\ m_h \stackrel{>}{\sim} 115\
{\rm GeV}.
\label{rel} 
\end{eqnarray}
Thus, we effectively have a mass relation between the light and heavy 
Higgs bosons  in the models. Numerically, we obtain the approximate relations:
\begin{eqnarray}
m_H &\approx & \left({m_h \over 115\ {\rm GeV}}\right)^{1/2} \ 360 \ {\rm GeV} \qquad
{\rm Broken\ case\ (Eq.\ref{10})};
\nonumber \\
m_H &\approx &\left({m_h \over 115\ {\rm GeV}}\right)^{1/2}\ 600 \ {\rm GeV} \qquad 
{\rm Unbroken\ case\ (Eq.\ref{11})}.
\label{rel2}
\end{eqnarray}

Phenomenologically, the broken mirror symmetry case, Eq.(\ref{10}), mimics the SM. The light Higgs, $h$, couples in exactly the same way as does the SM Higgs field, while the
heavier states, $H$, 
couple to the mirror sector. In this case, the experimental lower limit 
on $m_h$ is the 
same as the SM limit of approximately 115 GeV. Also, the 
upper bound on $m_h$ inferred from precision electroweak measurements is 
$m_h < M_{EW} \approx 186$ \cite{pdg}. There is no difficulty in satisfying these constraints.

In the case of the unbroken mirror symmetry, Eq.(\ref{11}), the light 
Higgs field $h$,  
couples to {\it both} of the sectors, with coupling strength $1/\sqrt{2}$ compared with the SM Higgs.
Thus the limit from precision electroweak measurements in this model is given by Eq.(\ref{a10}) with $c^2_{\omega}=1/2$.  
However, for $M_{EW} \approx 186$ GeV, this bound is inconsistent with the relation, Eq.(\ref{rel}). We conclude that the Coleman-Weinberg mechanism is consistent with existing
phenomenological bounds only for the broken-P situation\footnote{Of course this conclusion 
is only for the minimal mirror model with two sectors. In this context it will be interesting to study the case of generalized mirror models with N-sectors \cite{Foot:2005rn}. However, we will leave this study for the future.}.   Below, we therefore discuss the broken-P case only.

The experimental lower bound on the mass of the SM Higgs
boson, $m_h>115$ GeV, can be translated into 
a lower 
bound on the coupling $\lambda_2(\Lambda)\equiv 2\lambda(\Lambda)$:
\begin{equation}
\lambda_2(\Lambda)>4.2\left(\frac{115~{\rm GeV}}{m_h}
\right)~,
\label{21}
\end{equation}   
 for the broken P case of Eq.(\ref{10}).
 To find the domain of validity of perturbation theory we look, as before, at 
renormalisation group equations:
\begin{equation}
4\pi^2\frac{d\lambda_1}{dt}=3\lambda_1^2+\lambda_2^2~,
\label{23}
\end{equation} 
\begin{equation}
4\pi^2\frac{d\lambda_2}{dt}=3\lambda_1\lambda_2+\frac{1}{2} \lambda_2^2~,
\label{24}
\end{equation} 
where $\frac{1}{2}\lambda_1=\lambda+\delta$. As $\lambda_2(\Lambda)$ is significantly larger than any other coupling in the SM we can approximately neglect top-quark and gauge boson contributions.   
These equations must be 
supplemented by the constraint equation (\ref{8}), which reads 
\begin{equation} 
\lambda_1(\Lambda)=0~,
\label{25}
\end{equation}  
for the case (\ref{10}). 
Numerically solving these equations, we find that the position, $\mu_L$, of the 
Landau pole is $\mu_L \approx 3 \times 10^2 \Lambda \approx 10^5$ GeV  
(for $m_h$ at the experimental limit of 115 GeV).
The perturbative domain thus extends confortably above 
the electroweak scale.

\section{Conclusion}

We have examined the idea that the standard model Higgs boson might 
be the pseudo-Goldstone boson of broken scale invariance. The simplest 
version of this idea is the original Coleman-Weinberg model. 
The problem there is that the mass of the Higgs is too small, and the 
spontaneous symmetry breaking can only occur if the top quark is also very light. 
However, if there is an extended Higgs sector, then the additional scalar 
degrees of freedom can compensate for the heavy top quark, and 
give a Higgs mass in excess of the experimental lower limit, currently around 115 GeV.

Specifically, we have considered two phenomenologically consistent models. 
The first involves the addition of one (or more) real scalar fields. The  
additional bosonic degrees of freedom can lead to phenomenologically successful 
electroweak symmetry breaking. An even more constrained symmetry breaking sector 
arises in the mirror model, which has a discrete symmetry interchanging 
the standard model Higgs with a mirror partner. The discrete symmetry 
eliminates one parameter in the Higgs potential, and leads to a consistent  
electroweak symmetry breaking with the same number of parameters as in the 
standard model case.

The proposed scale-invariant  models (and their generalizations) can be testable in upcoming LHC experiments in the case of non-zero $\cos\omega$. The light PGB Higgs interacts with  Standard Model particles with couplings reduced by  factor $\cos\omega$  relative to the corresponding couplings of the Standard Model Higgs boson. In addition heavier Higgs-like boson with a mass correlated with the mass of PGB Higgs (see Eq.(\ref{a9}) can be observed. The couplings  of this heavy boson with Standard model particles is suppressed by $\sin\omega$ and its total decay width will be dominated by the decay width into two PGB Higgses.  In the case of  $\cos\omega=0$, the difference between the PGB Higgs and the Standard Model Higgs is rooted in the scalar potential, and, most probably, LHC will not be capable to distinguish among them.  Future linear collider (e.g. the proposed ILC) should be able to study this case.

\section{Acknowledgements}

This work was supported by the Australian Research Council.

\end{document}